          [2001/04/25 1.1 (PWD)]
\documentclass[a4paper,twocolumn]{esapub} % European paper
\usepackage{natbib}
\usepackage{graphicx}
\usepackage[latin1]{inputenc}
\usepackage[T1]{fontenc}
\title{INTEGRAL,
XMM-Newton and Rossi-XTE Observations of the State Transition of
the X-ray Transient and Black Hole Candidate XTE~J1720-318}
\author{M. Cadolle Bel}
\author[1,2]{A. Goldwurm}
\author[1,3,4]{J. Rodriguez}
\author[1,2]{P. Goldoni}
\author{S. Corbel}
\author{P. Sizun}
\affil[1]{Service d'Astrophysique, DAPNIA/DSM/CEA - Saclay, 91191
Gif-sur-Yvette Cedex, France} \affil[2]{Fédération de Recherche
APC, 11 place M. Berthelot, 75231, France}\affil[3]{Integral
Science Data Center, Chemin d'Ecogia, 16, CH-1290 Versoix,
Switzerland} \affil[4]{CNRS FRE 2052, France}
\author[5]{A.N. Parmar}
\author[5]{E. Kuulkers}
\affil[5]{Research and Scientific Support Department, ESA, ESTEC,
NL-2200 AG Noordwijk, The Netherlands}
\author[6]{F. Capitanio}
\author[6]{M. Del Santo}
\author[6]{A. Tarana}
\author[6]{P. Ubertini}
\affil[6]{IASF-CNR, via del Fosso del Cavaliere 100, 00133 Roma,
Italy}
\author[7]{J.P. Roques}
\author[7]{L.~Bouchet}
\affil[7]{Centre d'Etude Spatiale des Rayonnements, CNRS, Toulouse
Cedex 4, France}
\author[8]{R. Farinelli}
\author[8]{F. Frontera}
\affil[8]{Physics Department, University of Ferrara, 44-100,
Ferrara, Italy}
\author[9]{N.J. Westergaard}
\affil[9]{Danish Space Reasearch Institute, DK-2100 Copenhagen 0,
Denmark}

\def\eg{{\it e.g.}}
\def\etal{{\it et~al.~}}
\def\ie{{\em i.e. }}
\def\x1720{XTE~J1720-318}
\def\nh{N$_{\mathrm{H}}$}
\def\kir{$\chi^2_{red}$}

\begin{document}
\maketitle
\begin{abstract}
We report the results of extensive high-energy observations of the
X-ray transient and black hole candidate \x1720 performed with
INTEGRAL, XMM-Newton and RXTE. The source, which underwent an
outburst in January 2003, was observed in February in a spectral
state dominated by a soft component with a weak high-energy tail.
The XMM-Newton data provided a high column density \nh~of
$1.2\times 10^{22}$~cm$^{-2}$ which suggests that the source lies
at the Galactic Center distance. The simultaneous RXTE and
INTEGRAL Target of Opportunity observations allowed us to measure
the weak and steep tail, typical of a black-hole binary in the
so-called High/Soft State.\\
We could follow the evolution of the source outburst over several
months using the INTEGRAL Galactic Center survey observations. The
source regained activity at the end of March: it showed a clear
transition towards a much harder state, and then decayed to a
quiescent state in summer. In the hard state, the source was
detected up to 200~keV with a typical power law index of
$\sim$~1.8 and a peak luminosity of 7.5$\times
10^{36}$~ergs~s$^{-1}$ in the 2-100 keV band, for an assumed
distance of 8 kpc. We conclude that \x1720 is indeed representative of the class of the black hole X-ray novae which populate our Galactic
bulge and we discuss its properties in the frame of the spectral
models used for transient black hole binaries.
\end{abstract}

\section{Introduction}
X-ray Novae (XN) are low mass X-ray binaries where a compact
object normally accretes at very low rate from a late type
companion star (Tanaka~\&~Shibazaki 1996). Although they are
usually in quiescent state (and therefore nearly undetectable),
they undergo bright X-ray outbursts, with recurrence times of
several years, which last several weeks or even months before the
source returns to quiescence again. Most of the XN are associated
to dynamically proven Black Holes (BH) and indeed the great majority of
the known 18 Black Hole Binaries (BHB) as well as of the 22 binary
Black Hole Candidates (BHC) are transients
(McClintock~\&~Remillard 2003). Because of large changes in the
effective accretion rates that occur during the XN outbursts and
the very hard spectra they usually display, these sources are very
useful to study accretion phenomena and radiation processes at
work in BH, and are primary targets for high-energy
instruments.\\Since XN probably follow the Galactic stellar
distribution, they are concentrated in the direction of the bulge
of our Galaxy (with a higher concentration towards the center).
The SIGMA gamma-ray telescope onboard the GRANAT satellite, and
later the hard X-ray instruments onboard CGRO, RXTE and Beppo-SAX indeed
discovered and studied several (about 10) BHC XN in the bulge.
INTEGRAL, the INTErnational Gamma-Ray Astrophysical Laboratory
(Winkler \etal 2003) is a European Space Agency observatory that
began its mission on 2002 October 17, carrying four instruments:
two main gamma-ray instruments, IBIS (Ubertini \etal 2003) and SPI
(Vedrenne \etal 2003), and two monitors, JEM-X (Lund \etal 2003)
and OMC (Mas-Hesse \etal 2003). The IBIS coded mask instrument is
characterised by a wide Field of View (FOV) of 29°$\times$29°
(9°$\times$9° fully coded), a point spread function of 12' FWHM
and a sensitivity over the energy range between 20 keV and 8 MeV.
Thanks to its instruments performances and to the survey program
specifically dedicated to the Galactic Center (GC) region,
INTEGRAL is expected to allow the detection and the study of BH XN at a large distance and at weaker flux levels than before.
\\
\x1720 was discovered on 2003 January 9 with the ASM monitor onboard RXTE as a transient source undergoing an
X-ray nova like outburst (Remillard \etal 2003). The source flux
increased to the maximum value of $\sim$430 mCrab in 2 days, and
then its flux started to decay slowly. Follow up observations of
the PCA array onboard RXTE, have shown the
presence of a 0.6~keV thermal component and a hard tail. The
spectral parameters and the source luminosity suggested a BH (Markwardt 2003) in a High/Soft State (HSS). Soon after, a
radio counterpart was identified with the VLA and ATCA radio
telescopes (Rupen \etal 2003; O'Brien \etal 2003), leading to the
estimate of the most precise position
$\alpha_{J2000}=17^\mathrm{h}19^\mathrm{m}58.985^\mathrm{s}$,
$\delta_{J2000}=-31^\circ45^\prime01.109^{\prime\prime}$~(uncertainty$\pm~
 0.25^{\prime\prime}$). The detection of its infrared counterpart
(Nagata \etal 2003) implies an extinction compatible with the
location of \x1720 at large distance.
\\
\x1720 was observed by XMM-Newton, RXTE and INTEGRAL in
February during dedicated Target of Opportunity (ToO)
observations. It was then observed by INTEGRAL during the surveys
of the GC region performed in March and April and again from
August to October 2003. We report in this paper the results based
on these observations, starting with the description of the
available data and of the analysis procedures employed (Section
2). We then report the analysis results in Section 3 and we will
discuss them in Section 4.

\section{Observations and Data Reduction}
\begin{figure}
\centering
\includegraphics[width=1\linewidth]{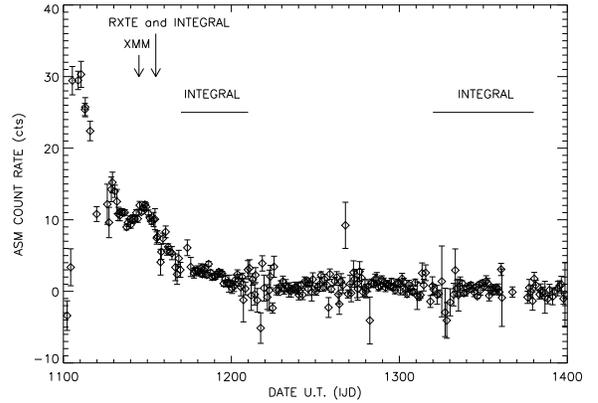}
\caption{RXTE/ASM light curve (2-12 keV) of the outburst of
\x1720. The arrows show the dates of the XMM-Newton, RXTE and
INTEGRAL simultaneous observations. The approximate periods of
later INTEGRAL observations are indicated by horizontal lines. The
universal time is in units of INTEGRAL Julian Days (IJD) where IJD
= MJD-51544 days.\label{LCLONG}}
\end{figure}

\x1720 was observed by XMM-Newton on 2003 February 20,
during a public 18.5 ks ToO. Preliminary analysis of these data
provided an improved X-ray position of the source
(Gonzalez-Riestra \etal 2003) confirming the association with the
radio and IR source. One week after, we performed an INTEGRAL ToO
observation of \x1720 which started on 2003 February 28 for a 176
ks exposure. The latter was conducted in coordination with a RXTE
ToO observation which lasted about 2~ks. The source was further
observed during the INTEGRAL Core Program during a series of
exposures dedicated to the GC survey, from March 25 to April 19
for a total of 551~ks observing time. Another 75 ks exposure on
the source has been accumulated during a ToO observation of
H~1743-322 (Parmar \etal 2003) in April 2003. The field containing
\x1720 has also been extensively monitored in the fall of~2003
during the second part of the 2003 INTEGRAL GC survey.
\\
The log of the observations and data used in this work is
summarized in Table~\ref{tab:log}. The ASM light curve (2-12 keV)
of~\x1720, showing the transient source outburst and the following
X-ray flux decay, is presented in Figure~\ref{LCLONG}. We also
indicate the sequence of the observations discussed in this work.

\begin{table*}[htbp]
\begin{center}
\caption{\label{tab:log}Log of the \x1720 observations analysed in this paper.
}
%\vspace{1em}
\begin{tabular}[h]{lllll}
\hline \hline
Spacecraft & Observation Date & Total & Instrument & Observation type\\
& (\# revolution) & exposure time &used& /Mode\\
\\
XMM-Newton & 02/20 & 18.5 ks & EPIC-PN & ToO/Small Window\\
RXTE & 02/28 & 2 ks & PCA+HEXTE& ToO\\
INTEGRAL & 02/28-03/02 (46)  & 176 ks & JEM X-2+IBIS & ToO~$^a$\\
INTEGRAL & 03/25-04/03 (54-57) & 361 ks & IBIS & GCDE \\
INTEGRAL & 04/06-04/07 (58) & 75 ks & IBIS & ToO on H 1743-322~$^a$\\
INTEGRAL & 04/12-04/19 (60-62) & 191 ks & IBIS & GCDE \\
INTEGRAL & 08/02-10/16 (97-122) & 605 ks & IBIS & GCDE \\
\hline
\end{tabular}
\end{center}
Note a: 5$\times$5 dithering pattern around the target.
\end{table*}

\subsection{XMM-Newton Data Analysis}
We present here only the data taken with the EPIC-PN camera on
board XMM-Newton. The PN camera was operating in Small Window
mode. We processed the data using the Scientific Analysis System
v.5.4.1 and the calibration files updated at the end of March
2003. We first filtered our data for background flares. Since
\x1720 was bright at the date of the observation (resulting in a
strong pile up in the PN camera), we adopted the selection
criteria suggested by Guainazzi (2001) to obtain the source
spectrum. We extracted the events from an annulus with an internal
radius of 15$^{\prime\prime}$, and an outer radius of
29$^{\prime\prime}$ around the position of \x1720. As we only used
single events, the effective exposure time of the extracted
spectrum was about 6~ks.\\
We obtained the background spectrum from a
sky region far from the source and we built the response matrix
(RMF) and ancillary response (ARF) files consistent with the
selections. The resultant spectrum was then fitted with
XSPEC v11.3.0 (Arnaud 1996) between 0.5 and 12 keV.\\

\subsection{RXTE Data Analysis}
We reduced and analysed the RXTE data with the LHEASOFT package
v5.2. We reduced the data from PCA and HEXTE following the
standard ways explained in the ABC of RXTE and the cook book. The
good time intervals (GTI) were defined when the satellite
elevation was $>$ 10$^\circ$ above the Earth limb, and the offset
pointing $<$0.02$^\circ$. We also chose to retain the data taken
when most of the Proportional Counter Units (PCU) were turned on
(a maximum of 5 here). We extracted the spectra from the standard
two data groups, from the top layer of each PCU. Background
spectra were produced with pcabackest  v3.0, using the latest
calibration files available for bright sources. RMF and ARF were
generated with pcarsp v8.0. Due to uncertainties in the PCA RMF,
we included some systematic errors in the spectra. To estimate the
level of those systematics, we reduced and analysed a
contemporary Crab observation. To obtain a reduced $\chi^2$ of
1 when fitting the Crab spectra, we set the level of systematics
as follows: 0.6$\%$ between 2 and 8 keV and 0.4$\%$ above 8 keV.\\
%\indent  We extracted 16 s resolution PCA light curves from
%standard 2 data using all PCUs and all layers, between 2 and 20
%keV (absolute channels 5-49), and corrected them for background.
We extracted source and background spectra for both
clusters (0 and 1) of HEXTE from the archive mode data, after
separating on and off-source pointings. We corrected the spectra
for dead-time, and produced the RMF and ARF with hxtrsp v3.1. Due
to dubious spectral information, we avoided detector 2 of cluster
1 in the spectral extraction.\\
We fitted the spectra
between 3-25 keV for PCA and 20-40 keV for HEXTE, due to poor
statistics in the HEXTE high spectral bands (detection at a level greater
than 3$\sigma$ is achieved only up to 30 keV).\\

\subsection{INTEGRAL Data Analysis}
An INTEGRAL observation is made of several pointings (science
windows, hereafter SCW) each having $\sim2200$~s exposure,
following a special pattern on the plane of the sky (Courvoisier
\etal 2003). Except for the $5 \times 5$ dithering mode for
revolutions 46 and 58, the entire GC region was observed in the
framework of the Galactic Center Deep Exposure (GCDE) program
(Winkler 2001). Deep exposures in the GC radian ($\pm30$~deg in
longitude,~$\pm20$~deg in latitude centered at \emph{l}=0,
\emph{b}=0) are obtained with a set of individual pointings
lasting
30 min each on a regular pointing grid.\\
All the INTEGRAL instruments are operating simultaneously. We
describe here mainly results obtained from the data recorded with
the ISGRI detector (Lebrun \etal 2003) of the IBIS telescope
covering the spectral range from 20 to 800 keV. For the first
observation set, when the source was very soft, we also present
data from the JEM-X instrument (3-25 keV). More complete results
from the JEM-X and the SPI data will be presented elsewhere (Cadolle Bel \etal 2004, accepted for publication in A\&A). The
IBIS data have been reduced with the Offline Scientific Analysis
(OSA) v3.0, to produce images and extract spectra for each SCW
(Goldwurm \etal 2003). We selected SCW for which the source was
within 8$^\circ$ from the telescope axis. For the spectral
analysis, we used a 10 linearly rebinned channel RMF and a
recently preliminary corrected ARF on the Crab. The resultant
spectrum was fitted between 20 and 200 keV, since above 200 keV
the source is not significantly detected and below 20 keV
systematic uncertainties are still very high. Systematics errors
of $10\%$ were applied to account for the residual effects of the
response matrix. For the image analysis, the background derived
from empty fields was subtracted before deconvolution and we used
a catalog of about 41 sources to analyse the images. The total
amount of IBIS data we processed was equivalent to about 1400 ks
of exposure time, however due to selections performed and the fact
that the source was very often off-axis, the
effective exposure time is reduced to 522 ks.\\
We reduced the JEM-X data with the latest available
software version OSA v3.0. Only the JEM-X2 monitor was turned on
during our observation. We selected the SCW were the source was
closest to the center of the field of view, from which we
extracted the spectra for an effective exposure time of 21 ks. We
fitted the resultant averaged spectrum between 3 and 26.5 keV,
with the standard RMF and ARF.
\section{Results}
\subsection{The High Soft State}
The XMM and INTEGRAL/RXTE observations of February 2003 caught
the source in a very soft state. The source appeared bright at low
energies, with a daily-averaged flux between 100 and 140~mCrab in
the 2-12 keV band. The JEM-X and PCA instruments detected the
source at very high significance and we could derive significant
spectra up to 20 keV. On the other hand, the high-energy emission
was quite weak. IBIS only marginally detected the source at a
level of 0.4 cts s$^{-1}$ or 1.9 mCrab in the 20-60 keV band,
providing only few data points at energies higher than 20 keV.
RXTE/HEXTE also provided low significant data points at energy
$\geq$ 20 keV. The individual and combined spectra are described
in section 3.3.

\subsection{The Transition to Hard State}
\begin{figure}[h!]
\centering
\includegraphics[width=0.7\linewidth,angle=270]{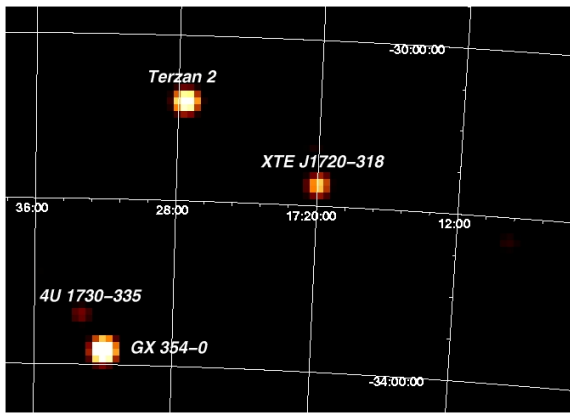}
\caption{The IBIS/ISGRI reconstructed sky image of the region
around \x1720 in the 20-40 keV band, using data from revolution 46
to 64. \x1720 appears at a significance level of 35$\sigma$ over
the background.\label{Mosa}}
\end{figure}

\begin{figure}
\centering
\includegraphics[width=0.7\linewidth,angle=270]{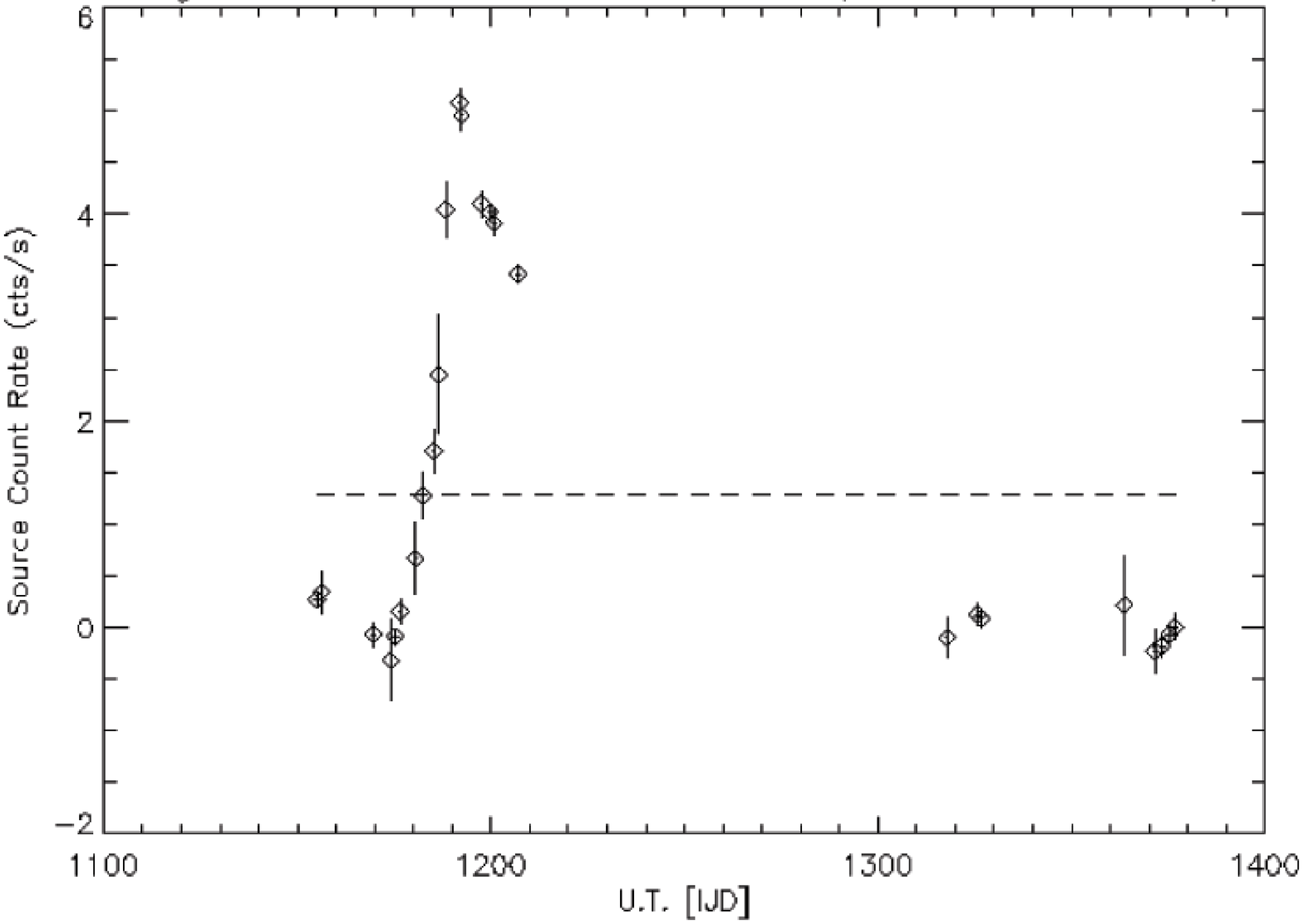}
\caption{The IBIS/ISGRI light curve of  \x1720 in the 20-60 keV
band, with time bins of 2 days, from revolution 46 to 122. The
dashed line represents the average flux over the whole period.
\label{LCTOT}}
\end{figure}

\begin{figure}
\centering
\includegraphics[width=0.7\linewidth,angle=270]{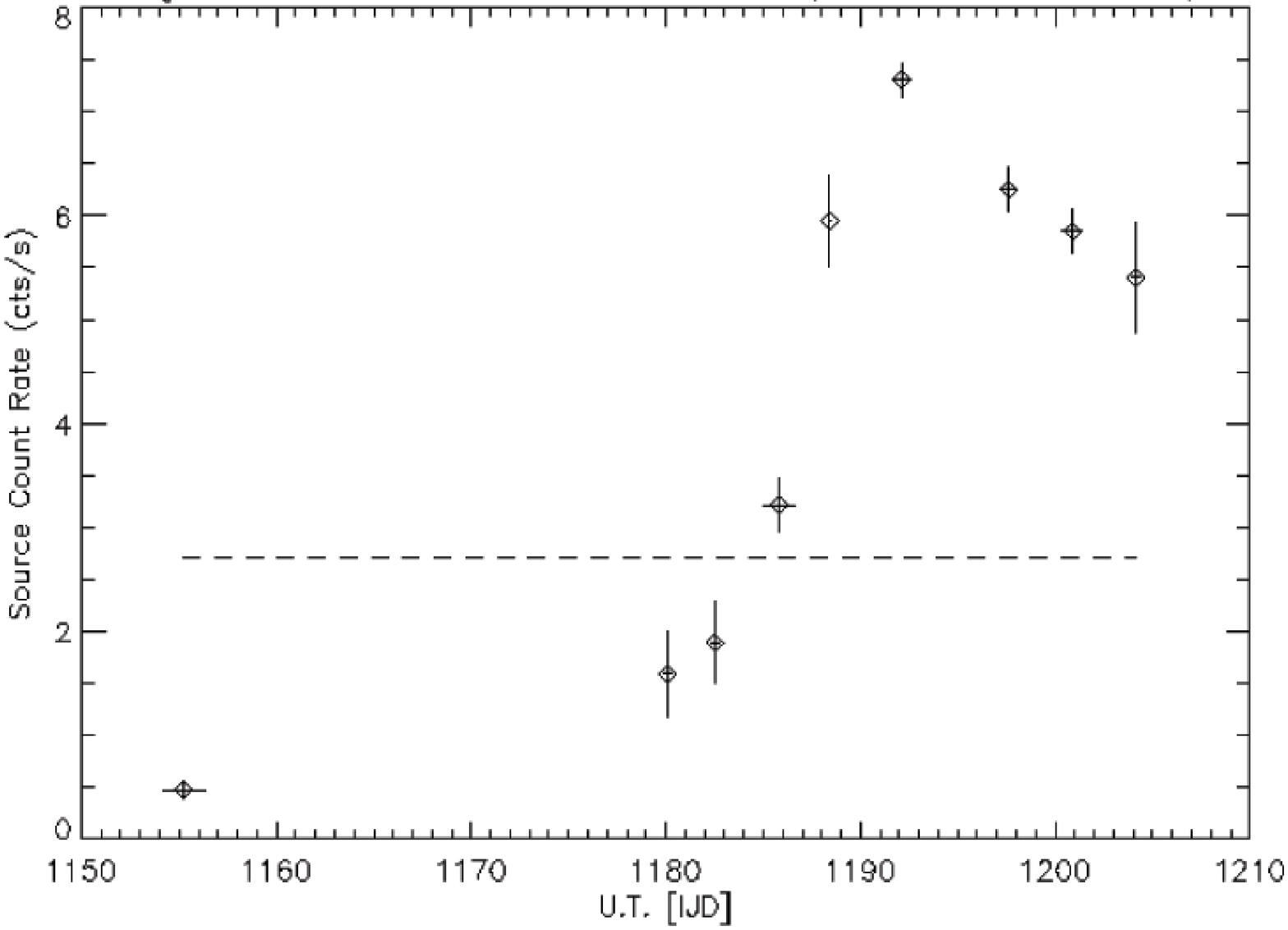}
\caption{The IBIS/ISGRI light curve of \x1720 in the 20-200 keV
band during the hard flare (from revolution 46 to 62), with time
bins grouping each revolution separately. The dashed line
represents the average flux over the whole period.\label{LCRED}}
\end{figure}

\begin{figure}
\centering
\includegraphics[width=0.75\linewidth,angle=270]{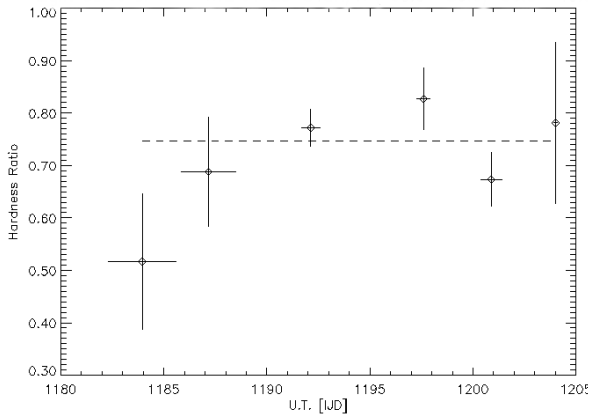}
\caption{IBIS/ISGRI source hardness ratio, defined as the ratio
between the source count rate in the 40-80 keV band and in the
20-40 keV band, during the \x1720 flare (revolutions 55 to 62).
The time bins are 3.5 days and the dashed line represents the
average hardness ratio.\label{HR}}
\end{figure}

Starting from March 25 (IJD=1180, revolution 54), the source
appeared to brighten in the INTEGRAL energy band. Since a similar
behavior was not seen in the ASM light curve, Goldoni \etal (2003)
proposed that the source was entering in a hard state.
\\
In the combined IBIS/ISGRI images obtained from the data of
revolutions 46 to 64 (see Figure~\ref{Mosa}), \x1720 is detected
at 35 $\sigma$ in the 20-40 keV range. The best position found
with IBIS from the 20-40 keV image is
$\alpha_{J2000}=17^\mathrm{h}20^\mathrm{m}01^\mathrm{s}$,
$\delta_{J2000}=-31^\circ45^\prime18^{\prime\prime}$ with an
accuracy of $0.9^\prime$ at $90\%$ of confidence level (Gros \etal
2003). This position is consistent with the most precise position
of \x1720 derived from radio data.%since the offset is only $1.8^{\prime\prime}$.
The high-energy source is therefore unambiguously associated to
the transient. We derived the source light curve in different
energy bands.
The IBIS light curve of \x1720 covering the whole year 2003 is
shown in Figure~\ref{LCTOT}. It was marginally detected during the
ToO observation of February 28 (revolution 46, IJD=1155). The
source decreased below the detection level when re-observed with
INTEGRAL about 10 days later, between March 9 and 20. \x1720
became visible again above 20 keV after March 25 (IJD=1180,
revolution 56) as shown on Figure~\ref{LCRED} where we focus on
the flare period: its 20-200 keV flux, was then around 3.3
cts~s$^{-1}$ or 15.5~mCrab and increased to a maximum level of
7.4 cts~s$^{-1}$ ($\sim$~37.5 mCrab) on April 6 (revolution 58,
IJD=1192). After this, the flux gradually decreased to the value
of 5.5 cts s$^{-1}$ (revolution 62). When the INTEGRAL GC survey
included the source again in the IBIS field of view in mid-August
2003, the transient was not detected and remained under the
detection level for the rest of 2003.\\Figure~\ref{HR} reports the
hardness ratio (HR) measured during the observed increase in the
high-energy source flux. There is no significant variation in the
HR around its mean value of 0.75, only a slight indication of a
softer HR ($\sim$~0.5) at the beginning of the flare. We therefore
used the whole data of this hard flare to build up an average
spectrum.

\subsection{Spectral Results}
\subsubsection{The High/Soft State Spectrum}
\begin{figure}
\centering
\includegraphics[width=0.7\linewidth,angle=270]{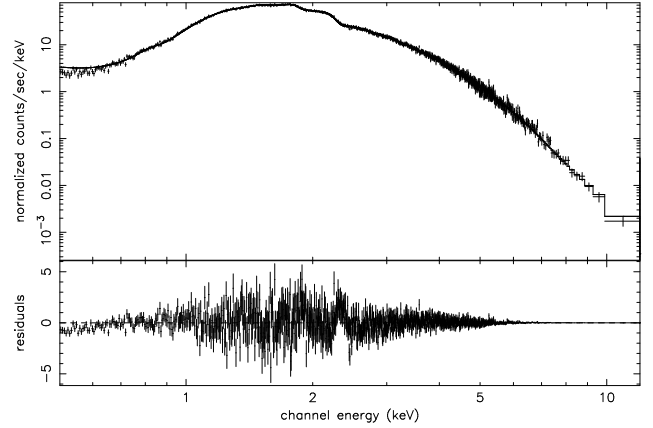}
\caption{XMM-Newton/EPIC-PN spectrum of \x1720. The best-fit
model, an absorbed MCD plus a power law, is over
plotted as a solid line. Residuals (in counts s$^{-1}$ keV$^{-1}$)
are also shown.\label{XMM}}
\end{figure}

\begin{figure}
\centering
\includegraphics[width=1.\linewidth]{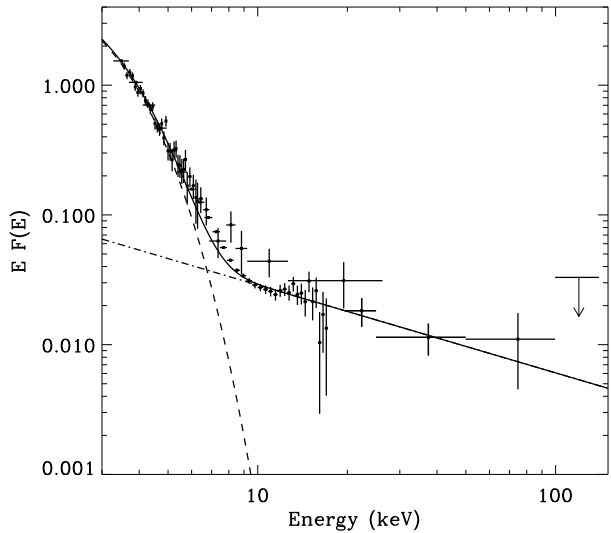}
%{spec_fitname.ps}
\caption{Unabsorbed EF(E) spectrum of \x1720 (units of keV cm$^{-2}$ s$^{-1}$) along with the best-fit model MCD plus powerlaw. {\it Dashed}: MCD. {\it Dotted dashed}: power law. {\it Thick}: total model. \label{SIMULT}}
%\caption{Joint RXTE/PCA, RXTE/HEXTE, INTEGRAL/JEMX-2 and
%INTEGRAL/IBIS spectra of \x1720 during the observations of end
%February 2003. The best-fit model, an absorbed black-body disc
%plus a power law, is over plotted as a solid line to the
%data.\label{SIMULT}}
\end{figure}

We have fitted the XMM-Newton EPIC-PN data with a model
composed of an absorbed multi-colour black-body disc (MCD, Shakura \& Sunyaev 1973 and Mitsuda
\etal 1984) plus a power law. A single absorbed MCD model
leads to a poor fit, as does a single absorbed power law.
The best-fit parameters derived from our analysis are given in
Table~\ref{tab:XMM-RXTE-INT}. We have found for \nh~the value of
(1.24$\pm0.02)\times10^{22}$~cm$^{-2}$. The unabsorbed flux in the
0.5-10 keV range was $7.31\times10^{-9}$~ergs~cm$^{-2}$~s$^{-1}$.
Assuming a distance of 8 kpc (see discussion), the luminosity in
the 0.5-10 keV range is then $5.57~\times~10^{37}$~ergs~s$^{-1}$.
The disc component accounts for $82\%$ of the total luminosity. If
we assume a line of sight inclination angle of 60°, we find an internal disc radius of
48.9~$\pm~0.4$~km. The spectrum is shown on Figure~\ref{XMM}.\\
\begin{table*}[htbp]
\begin{center}
\caption{\x1720 best-fit spectral parameters for the XMM-Newton
ToO, for the simultaneous RXTE/INTEGRAL ToOs of February and for
the INTEGRAL detected hard flare (revolutions 55 to 62), with
their $90\%$ confidence level errors.}\vspace{1em}
\begin{tabular}{ccccccc}
\hline \hline
Instrument&Date&Disc Tempe-&Disc&Photon&\kir&Flux~$^b$\\
&(U.T.)&rature(keV)&Radius(km)~$^a$&Index&(dof)&$\times$10$^{-9}$ergs s$^{-1}$ cm$^{-2}$\\
\\
XMM-Newton & 02/20&  0.67$\pm0.01$ &48.9$\pm0.4$&$2.69_{-0.57}^{+0.44}$ & 1.2(1102)& 2.36\\
RXTE+INTEGRAL & 02/28-03/02&0.59$\pm0.01$&84$\pm4$&$2.63_{-0.22}^{+0.34}$ &  0.9(94) & 3.26\\
INTEGRAL & 03/27-04/19&-&-&1.84$\pm0.11$ & 1.6(7) & 0.98\\
\hline
\end{tabular}
\label{tab:XMM-RXTE-INT}
\end{center}
Notes:
a: Disc radius R in units of km is given by K~$=~(\frac{R}{D})^{2}\times\cos{\theta}$
where K is the disc normalisation, D is
the distance to the source in units of 10 kpc and $\theta$ the
inclination angle of the disc.\\
b: Unabsorbed 2--100 keV flux.
\end{table*}
We have applied the same absorbed MCD plus power law model to a simultaneous fit of the RXTE/PCA,
RXTE/HEXTE, INTEGRAL/JEM-X and INTEGRAL/IBIS data taken about 8
days later and we obtained the best-fit parameters reported in
Table~\ref{tab:XMM-RXTE-INT}. To account for uncertainties in
relative instruments calibrations, we let a multiplicative
constant to vary in the fit of the different data sets. Taking the
RXTE/PCA spectrum as the reference, the derived constants are all
found very close to 1 for each instrument, except for RXTE/HEXTE
for which we got a factor of 0.7. This is compatible with the
level of cross-calibration normalization between the two RXTE
instruments. As RXTE and JEM-X are not suited to determine
interstellar absorption (energy lower boundary is $\sim$~3~keV),
we fixed the \nh~to the value obtained from the XMM-Newton fits.
We also added a gaussian line at the iron fluorescent line
energies to account for a feature present in the RXTE data. The
line centroid was found to be $6.45_{-0.35}^{+0.16}$ keV with an
equivalent width of $572_{-178}^{+307}$~eV. However, this line was
not present in the data obtained with XMM-Newton. To check the
reality of this line, we reperformed the fit of the EPIC PN
spectrum by adding to the best fit continuum model an iron line at
a fixed energy and width equal to the ones found from the RXTE
data. We obtained an upper limit for such a line of 74.4~eV
equivalent width at the 90$\%$ confidence level. This upper limit
suggests that the line seen with RXTE is probably due to an
incorrect background substraction and not to \x1720. For example,
it may be due to contamination by the Galactic ridge emission
(Revnivtsev 2003). For this reason, we did not included the line
for the fit of the JEM-X data. In spite of the low significance
level of the detection, the IBIS/ISGRI data allow us to study the
source up to higher energies than with RXTE/HEXTE alone because of
the higher sensitivity of ISGRI and the longer exposure time.
The derived spectrum is shown on Figure~\ref{SIMULT}.\\
According to the value of the photon-index, we found that
the source was clearly in the HSS, where the thermal component
from the disc dominates. The internal radius is now given by
84~$\pm~4$~km and the disc flux contribution around 93$\%$ of the
total luminosity in the 2-100 keV range. Indeed, there is a slight
evolution between the XMM-Newton parameters (internal radius and
temperature) and the same parameters found one week after by RXTE
and INTEGRAL. But all these data taken during the last week of
February are consistent with the BH XN \x1720 being in the
HSS.
\subsubsection{The Low/Hard State Spectrum}
\begin{figure}
\centering
\includegraphics[width=0.7\linewidth,angle=270,keepaspectratio]
{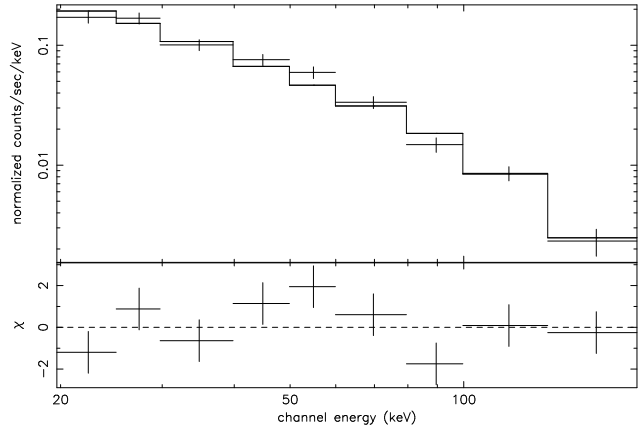} \caption{IBIS/ISGRI (20-200 keV)
spectrum of \x1720 using data from the end of March to mid-April. The
best-fit power law model is overplotted as a solid line. Residuals
(in $\sigma$ units) are also shown. \label{POW}}
\end{figure}

As discussed above, the IBIS data from revolutions 55 to 62 are
consistent (\ie no variation of HR) and can be summed to derive
the average spectrum reported in Figure~\ref{POW}. We fitted this
spectrum with a simple power law model between 20 and 200 keV.
Above 200 keV the source is not significantly detected. Again,
$10\%$ systematics were applied. The best-fit photon-index
returned from the fits is 1.8 (see Table~\ref{tab:XMM-RXTE-INT}),
which reveals that the spectrum of \x1720 is much harder than
observed in February. In addition to the power law model, we have
tried to fit the data set with a comptonisation model (Sunyaev \&
Titarchuck 1980). The derived parameters are $49 _{-20}^{+51}$~keV
for the temperature and $2.6 _{-1.2}^{+1.4}$ for the optical
depth, with a reduced $\chi^2$ of 1.8 (6 degrees of freedom). The
$\chi^2$ is not significantly better than the one obtained with
the single power law and our results do not allow to choose
between the two models, \ie it seems that no high-energy cut off
is clearly detected. On the other hand, the derived thermal
comptonisation parameters are very much consistent with those
found in BHB in the so-called Low/Hard State
(LHS).

\section{Discussion}
The high equivalent absorption column density derived from the
XMM-Newton data suggests that \x1720 lies at the GC distance or
even further. This would place the source in the Galactic bulge
and we will, therefore, assume a source distance of 8 kpc. When
observed with XMM-Newton, about 40 days after the outburst peak,
\x1720 was in a HSS, characterized by a strong soft (thermal)
component, well modeled by a disc emission model with an
inner disc temperature of kT~$\sim$~0.6 keV, and a weak power law
tail. The source was found in HSS also at the end of February,
when we could measure, with higher precision with INTEGRAL and
RXTE, the steep power law slope index of 2.6. In both
observations, the disc component accounted for more than 80$\%$ of
the 2-100 keV source luminosity, estimated at the end of February
at 2.5~$\times~10^{37}$~ergs~s$^{-1}$. The source did not change
state during the decay phase which started after the outburst peak
and lasted till about mid-March, although slightly different
spectral parameters of the soft component were measured during the
INTEGRAL/RXTE observations (\ie a lower temperature and a larger
inner-disc radius). This could indicate that the cool accretion disc
was receding from the BHC, in agreement with certain
interpretation of the outburst evolution in XN, but could also be
linked to a specific spectral variation of a secondary flare.
Indeed the XMM-Newton observation took place during a weak
secondary peak which occurred in the decay phase (see
Figure~\ref{LCLONG}), which was also observed in infrared (Nagata
\etal 2003).
\\
A more dramatic change in the source behaviour was observed with
INTEGRAL at the end of March. We observed a soft to hard spectral
state transition of the X-ray transient about 80 days from the
outburst peak. The luminosity increased in about 10 days from
below the INTEGRAL detection level to an extrapolated 2-100 keV
luminosity of $\sim$ 7.5$\times 10^{36}$~ergs~s$^{-1}$, without
any similar increase in the low-energy flux measured by RXTE/ASM.
The spectrum was hard and well described by a power law of index
1.8 or a thermal comptonisation model with a (weakly constrained)
plasma temperature of 49 keV and an optical thickness of 2.6. No
clear spectral break was observed.\\
The high peak luminosity, the fast rise and slow decay time
scales, the high soft spectral state and the late transition to a
LHS with spectral parameters typically observed in other
(dynamically confirmed) BH transients, like \eg~XTE~J1550-564
(Sobczak \etal 2000; Rodriguez \etal 2003), GRO~J1655-40
(Sobczak \etal 1999, see also McClintock \& Remillard 2003) or XN Muscae 1991/GRS 1124-68 (Goldwurm \etal1992, Grebenev \etal 1992 and Ebisawa \etal 1994) show that XN \x1720 is probably a new XN and BHC of the
Galactic bulge.
\\
Although there is little doubt about the origin of the
soft thermal component and its modeling, the interpretation of
the high-energy tail and its connection to the spectral states
remain rather controversial. In the HSS, most of the X-rays are
radiated by the accretion disc. The decay phase of XN in the HSS
is clearly linked to the decrease of the effective accretion rate.
The standard Shakura \& Sunyaev (1973) $\alpha$-disc, however,
cannot produce hard radiation (in either of the spectral states).
In the LHS, the hard component is generally attributed to thermal
comptonisation of the disc soft radiation by a hot plasma (Sunyaev
\& Titarchuk 1980, Titarchuk 1994) located above the disc or in
the inner part of the system, around and very close to the BH.
However, the details of the geometry and of radiation mechanisms
at work are still not understood; the processes which lead to the
spectral transition and the possible role of non-thermal
(synchrotron) radiation are still very uncertain. For example, one set of models
which explain the above geometry and the comptonisation origin of
the hard emission in LHS are those based on Advection Dominated
Accretion Flows (ADAF). %The latter was
%invoked to account for the high-energy tail (steeper and without
%high-energy break) observed during the HSS.
Alternatively,
comptonisation on a population of (thermalised) electrons with
bulk motion (\eg~Titarchuk \etal 1997) may be responsible for the presence of the high-energy tail in HSS or in intermediate states. The detection and study of the XN of the Galactic bulge
with INTEGRAL will possibly provide more data on this kind of
objects and will thus improve our understanding of the physics of
BHB.
\\
Thanks to the high sensitivity of INTEGRAL, it has been
possible to study a faint source in the Galactic bulge, to detect
a spectral transition confirming the probable BH nature of the
object and to obtain a spectrum up to 200 keV. Further analysis of
the INTEGRAL data of \x1720 are in progress and are reported
in Cadolle Bel \etal 2004 (accepted for publication in A\&A).

\section*{Acknowledgements}
MCB thanks J. Paul and P. Ferrando for careful reading and
commenting the manuscript. JR acknowledges financial support from
the French Space Agency (CNES). The present work is based on
observations with INTEGRAL, an ESA mission with instruments and
science data center funded by ESA member states (especially the PI
countries: Denmark, France, Germany, Italy, Switzerland, Spain,
Czech Republic and Poland, and with the participation of Russia
and the USA) and with XMM-Newton, an ESA science mission with
instruments and contributions directly funded by ESA member states
and the USA (NASA).

\end{document}